\documentclass[10pt,twocolumn,amsmath,amssymb,superscriptaddress,noshowpacs]{revtex4}

\usepackage{graphicx}
\usepackage{dcolumn}
\usepackage{bm}

\begin{document}

\title{Spherically Symmetric Inflation\footnote{This article was submitted by the authors in English.}}

\author{R. S. Perez}
\email[]{rperez@cbpf.br}
\affiliation{Institute of Cosmology, Relativity and Astrophysics, CBPF, R. Xavier Sigaud 150, Urca 22290-180 Rio de Janeiro, Brazil}

\author{N. Pinto-Neto}
\email[]{nelsonpn@cbpf.br}
\affiliation{Institute of Cosmology, Relativity and Astrophysics, CBPF, R. Xavier Sigaud 150, Urca 22290-180 Rio de Janeiro, Brazil}

\begin{abstract}
\noindent {\bf Abstract $-$ }It is shown in this letter that in the framework of an inhomogeneous geometry and 
a massive non self-interacting scalar field with spherical symmetry, one needs
a homogeneous patch bigger than a dizaine of horizons in order to start inflation.
The results are completly independent of initial conditions on the spatial distribution
of the scalar field. The initial condition on the metric parameters are also justified.
This is a generalization of the results obtained in Ref.~\cite{calzetta}, showing that their
conclusions are rather robust.
\end{abstract}

\maketitle

\section{Introduction}

The inflation paradigm \cite{inflation} is an attempt to solve some of the drawbacks of the Standard
Cosmological Model, described by the homogeneous and isotropic Friedmann-Lema\^itre-Robertson-Walker 
(FLRW) geometries,

\begin{equation}
\label{m}
ds^2 = - {\rm d}t^2 + \frac{{a(t)}^2}{1 + \frac{k}{4}r^2}[{\rm
d}r^2 + r^2({\rm d} \theta ^2 + \sin ^2 (\theta) {\rm d} \varphi ^2)] \quad,
\end{equation}
where the curvature $k$ of the spacelike hypersurfaces can take the values $0$, $1$,$-1$.

These problems are related to the very special initial conditions associated with this model.
The first one is called the flatness problem: the present energy density of the Universe is observed
to be very close to the critical density, and it must have been much closer in the past, implying
that the observed homogeneous and isotropic space must be almost flat.
This problem comes from the classical equation
giving the density $\rho(t)$ relative to the critical one, $\rho_{c}(t) = 3 H^2(t) /(8\pi G)$, where $H\equiv \dot{a}/a$ is the Hubble expansion
rate, which reads
\begin{equation}
\frac{d |\Omega-1|}{d t}  = -2 \frac{\ddot{a}}{\dot a^{3}},
\label{Omega1}
\end{equation}
where $\Omega \equiv\rho/\rho_{c}$.
As $\Omega$ is close to unity now, 
Eq. (\ref{Omega1}) implies that it
must have been arbitrarily closer to unity in the past in the usual Friedmann model,
where one has a decelerated ($\ddot a <0$) expansion ($\dot a >0$) since the
initial singularity, and hence
$|\Omega-1|$ must have been an ever increasing function of time. In other words, $\Omega =1$ is an unstable point
in this model, and to have our old Universe still near this point implies an incredibly fine tuned value for 
$\Omega$ around the Plank era, $\Omega = 1 \pm 10^{-50}$.

The second problem is much more involved, and concerns the use of the very special homogeneous and isotropic geometry (\ref{m})
in order to describe our Universe. Either one has in hands a theory of initial conditions (perhaps quantum cosmology) which
selects this particular geometry from the possible infinite many inhomogenous solutions of the Einstein's equations, or there
was a physical mechanism of homogeneization and isotropization which has taken place in the early Universe leading the spacetime
geometry to the form of Eq.~(\ref{m}). In this paper, we will be concerned with the second approach.

The first basic assumption one must take for a physical homogeneization and isotropization of the Universe is the requirement that
its observed parts had some causal contact sometime in the past. However, if the Universe had a begining some
$14$ billion years ago, it can be shown that by the time of recombination, when the cosmic background radiation began
to propagate independently from matter, one had approximately $100$ regions without causal contact presenting
the cosmic radiation already with the same temperature. This is the so called horizon problem. Only after taking
care of this issue, one can begin to think about a physical mechanism of homogeneization and isotropization.

The flatness and horizon problems can be solved by inflation. It consists of the idea that the early Universe experienced
a brief but violent accelerated expansion, which turned $\Omega$ very close to one at this time (see Eq.~(\ref{Omega1}), 
and sufficiently enlarged a small causally connected piece of the Universe to its observed size. Furthermore, it induces
a mechanism for the origin of matter fluctuations, which gave rise to structure formation \cite{muk}.

However, it seems that the homogeneity problem is not solved by inflation. In fact, there are works showing
that in order for inflation to start in some region, one needs a homogeneous patch of a few horizons size at this region \cite{calzetta,piran}.
In order to circumvent this problem, some people evoque the Anthropic Principle \cite{ant}: in the regions where inflation does not
happen, it is not possible to have gallaxies, stars, and hence, intelligent life. Nevertheless, if one wants to rely on this Principle,
one could argue in the same way to justify the FLRW geometries (\ref{m}) without necessity of any period of inflation:
the difference would be the need of a much bigger homogeneous region containing a huge number of horizons to begin with,
with the apropriate initial perturbations printed on it. However, in an infinite inhomogeneous Universe, at least one of
such regions would exist, and the Anthropic Principle would select it. Without the Anthropic Principle, the role of inflation 
would then be to reduce drastically the size of the initial homogeneous region, increasing the number
of possible regions which can behave like our Universe. In that case, one would then need a more precise knowledge
of the possible cosmological scenarios after the Planck era, and a measure of such regions, which would lead us back to
the first approach to solve the homogeneity problem: a theory of initial conditions. In such a situation, inflation could alleviate 
but it would not solve alone this issue.

The homogeneity problem in the inflationary scenario is usually investigated numerically 
\cite{piran,numerico1,numerico2,numerico3}, but there are some
analytical studies in the literature \cite{analitico1,analitico2,causalidade1,causalidade2}. 
One of them \cite{calzetta}, deals with an inhomogeneous geometry and 
a massive non self-interacting scalar field with spherical symmetry in order to derive the size of the homogeneous patch 
necessary to yield sufficient inflation, under the assumption of certain initial conditions for the geometry and
scalar field. In this letter, we show that the results obtained in Ref.~\cite{calzetta} are independent of the choice
of initial conditions for the scalar field, and we try to justify some of the choices related to the spacetime geometry.
Hence, the result implying the necessity of a homogeneous patch bigger than a dizaine of horizons in 
order to start inflation is rather robust in this framework.

Note that we never make any splitting of the geometry on a homogeneous
background and small perturbations around it. We deal with full general
relativity and all their non-linear equations: deviations from homogeneity are
not considered to be small. This is because we are interested on the
homogeneity problem itself, and restricting ourselves to fluctuations on a
homogeneous geometry is not satisfactory as one would be assuming homogeneity
from the beginning. The treatment of small linear fluctuations around
inflationary models is made in many other papers in order to study the
evolution of structures and the back-reaction problem once inflation has
started. Here we are interested in the more basic question concerning how
inflation itself begins in a spherically symmetric inhomogeneous geometry.

In the next section we present the model, in section III we discuss the initial conditions, and in section IV
we present our results concerning inflation. We end this letter with conclusions and comments.

\section{The model}

The model consists of an asymptotically flat universe with open spatial sections and spherical symmetry. We write a spherically symmetric geometry in the following form:
\begin{equation}
ds^2=e^{2\alpha (\eta,r)}[-d\eta^2 + dr^2+e^{2\beta(\eta,r)}r^2d\Omega], \label{met}
\end{equation}
where $\eta$ is the time coordinate, and $\alpha$ and $\beta$ are arbitrary dimensionless functions of $\eta$ and $r$.
The matter content is a massive scalar field minimally coupled to gravity, whose energy-momentum tensor is given by
\begin{equation}
T_{\mu\nu} = \Phi_{,\mu}\Phi_{,\nu} - \frac{1}{2}g_{\mu\nu}[\Phi^{,\alpha}\Phi_{,\alpha} + m^2 \Phi^2]. \label{mom}
\end{equation}

We shall use units where $16\pi = G = 1$, with $\eta$ and $r$ having dimensions of length, while $m$ and $\Phi$ have dimensions of inverse length. From the expressions above, the Einstein's equations yield two dynamical equations and two constraint equations concerning the functions $\alpha$ and $\beta$. The constraint equations are the (00) and (01) components of the Einstein tensor, while the dyamical equations involve the (11) and (22) ones. The other components are trivial due to the spherical symmetry of the model, except for the equation involving the (33) component, which is equal to the (22) equation. Note that the right-hand-side of the following equations miss a length$^{2}$ factor due to our choice of units ($G=1$). The constraint equations are, therefore,
\begin{eqnarray}
&&3\dot{\alpha}^2+\dot{\beta}^2-4\dot{\alpha}\dot{\beta}+2\beta''-2\alpha''-\alpha'^2 -3\beta'^2+ \nonumber \\ 
&&~~~ +4\alpha'\beta'+6r^{-1}\beta'-4r^{-1}\alpha'-r^{-2}(1-e^{2\beta}) =\nonumber \\
 &&~~~~~~ =\frac{1}{4}(\dot{\Phi}^2+\Phi'^2+m^2e^{2\alpha}\Phi^2) \label{00}
\end{eqnarray}
and
\begin{eqnarray}
\dot{\alpha}'-r^{-1}\dot{\beta}-\dot{\beta}'-\dot{\alpha}\alpha'+\dot{\beta}\beta'=\frac{1}{4}\dot{\Phi}\Phi', \label{01}
\end{eqnarray}
where the primes and dots denote derivatives with respect to the radial and time coordinates, respectively. We will use combinations of the equations involving the (22) and (11) components as our dynamical equations. The first one is the difference between the (22) and (11) equations, namely
\begin{eqnarray}
r^{-2}(1&-&e^{2\beta})+2r^{-1}\alpha'+2\alpha'^2-2\alpha''+\nonumber \\ 
&&+\ddot{\beta}+\beta'' +2\dot{\alpha}\dot{\beta}-2\alpha'\beta'-2\dot{\beta}^2=\frac{1}{2}\Phi'^2,  \label{11} 
\end{eqnarray}
while the second is the difference between twice the (22) equation and the (11) one:
\begin{eqnarray} 
-2\ddot{\alpha}&+&4\alpha''-\dot{\alpha}^2+\alpha'^2-2r^{-1}\beta'-2\beta''+\dot{\beta}^2+\beta'^2 +\nonumber \\ 
&-&r^{-2}(1-e^{2\beta})=\frac{1}{4}(\dot{\Phi}^2-3\Phi'^2-m^2 e^{2\alpha}\Phi^2) \label{22}
\end{eqnarray}
To complete the set, the Klein-Gordon equation for the scalar field is given by
\begin{equation}
\ddot{\Phi}-\Phi''+2(\dot{\alpha}-\dot{\beta})\dot{\Phi}-2(\alpha'-\beta'-r^{-1})\Phi'+m^2e^{2\alpha}\Phi=0.
\end{equation}

\section{Initial Conditions}

We now check if this problem is well posed \cite{wald}, i.e., if we have a self-consistent Cauchy initial data. Among the most simple choices, we can set $\dot{\Phi}=0$ and $\beta=\dot{\beta}=0$. Note that the time evolution given by the Einstein-Klein-Gordon dynamics will subsequently modify these values.

The initial conditions $\beta=\dot{\beta}=0$ are in accordance with the
Weyl curvature hypothesis \cite{weyl}, which states that the Universe should be initially conformally flat.
This condition relies on the assumption that the gravitational entropy should be given by the Weyl tensor,
and stating that it was zero in the beginning should be equivalent to say that the Universe began in a state
of minimum entropy, as Boltzmann conjecture in order to justify the arrow of time. However, up to now,
this is just a speculation.

The choice of the initial condition $\dot{\Phi}=0$ at $\eta=\eta_0$ relies on the
fact that the  kinetic term of a scalar field decays exponentially fast in an
expanding geometrical patch, at least in the framework of chaotic inflation
(see Ref. \cite{piran}, page 267, and references therein). Furthermore, it seems to us
that considering $\dot{\Phi}\neq 0$ at $\eta=\eta_0$ will make things worst for the
occurrence of inflation, as long as it can happen only when the kinetic term is
negligible with respect to the potential term. Hence, the final result of our
paper should be considered as the minimum requirement for the occurrence of
inflation in the framework of spherically symmetric general relativity under
the Weyl curvature hypothesis.

Concerning the initial shape of the scalar field, we will try do keep this as general as possible, so we can write the following equation for $\Phi$ at time $\eta_0$
\begin{equation}
\Phi(\eta_0,r)=\Phi_0+f(r), \label{init}
\end{equation}
where $\Phi_0$ stands for the homogeneous part of the inflaton field and $f(r)$ is an arbitrary function of $r$ which is asymptotically null at spatial infinity. With this in hand, in the hypersurface $\eta=\eta_0$, equation (\ref{01}) reduces to
\begin{equation}
 \dot{\alpha}'-\dot{\alpha}\alpha'=0.
\end{equation}
The solution of this equation is $\dot{\alpha}=Ce^{\alpha}$. It can be easily shown that the constant $C$ is fixed by consistence requirements between the two constraint equations when we go far from the origin. Hence, we obtain
\begin{equation}
\dot{\alpha}=\frac{1}{2\sqrt{3}}m\Phi_0e^{\alpha}. \label{alfa1}
\end{equation}
Putting this result into equation (\ref{00}), we get the radial evolution for $\alpha$,
\begin{equation}
2\alpha''+\alpha'^2+4r^{-1}\alpha'=-\frac{1}{4}[\Phi'^2+m^2(\Phi^2-\Phi_0^2)e^{2\alpha}]. \label{alfa2}
\end{equation}
Near the spatial origin $r \rightarrow 0$, equation (\ref{alfa2}) has a power law solution, which is
\begin{equation}
\alpha \sim \bar{\alpha}-\frac{1}{24}\left[m^2\Phi_0 e^{2\bar{\alpha}} \bar{f} + \frac{1}{2}\left(\bar{f}'+\bar{f}^2\right)\right] r^2, \label {alfa3}
\end{equation}
where the bars above $f$ and $\alpha$ denote the values of these functions at $r=0$. Those are the expressions that we will use from now on. It can be shown \cite{calzetta} that equation (\ref{alfa2}) has an exact solution, and hence the Cauchy initial data is self-consistent.

\section{Conditions for inflation}

First of all, we remind the reader that there are conditions on the value of the homogeneous field related to the observations \cite{piran}. The fluctuations on the cosmic microwave background puts a limit on $\Phi_0$, which in our units is $m\Phi_0^2 < 10^{-4} \times 8\sqrt{3}$. On the other hand, the requirement of 70 e-folds of suficient inflation sets a minimum value to the scalar field, namely $\Phi_0 > \sqrt{561}$. Now we move to the conditions of our model.

From equation (\ref{11}), dropping the $\beta$ and $\dot{\beta}$ terms according to our choice of initial conditions, we get, near the origin,
\begin{equation}
\ddot{\beta}=\frac{1}{2}\Phi'^2 \sim \frac{1}{2}\bar{f}'^2, 
\end{equation}
and from this expression we can say that $\beta$ will be small for physical times around
\begin{equation}
\tau_\beta \sim \frac{e^{\alpha}}{\bar{f}'}.\label{tau}
\end{equation}

The first condition for inflation is that $\beta$ remains small during one e-fold, i.e., $\tau_\beta > H^{-1}$. Once inflation has started, any anisotropy will decay according to the ``no-hair'' conjecture, so $\beta$ will remain small for the rest of the process. We will see in the following lines that this condition is less restrictive than the others related to the function $\alpha$ of the metric.

From the evolution equation (\ref{22}) we have
\begin{equation}
2\ddot{\alpha}-4\alpha''+\dot{\alpha}^2+\alpha'^2=\frac{1}{4}(3\Phi'^2+m^2e^{2\alpha}\Phi^2).
\end{equation}
Using the radial solution (\ref{alfa3}) for $\alpha$, we get the dynamical equation
\begin{eqnarray}
8\ddot{\alpha} + 4\dot{\alpha}^2 &=& 3\bar{f}'^2 + 16\alpha''   +\nonumber \\
 &+& m^2(\Phi_0\bar{f}+ \bar{f}^2)e^{2\bar{\alpha}}+\Phi_0^2m^2e^{2\bar{\alpha}} \label{alphadyn}
\end{eqnarray}
because, according to this solution, the term $\alpha'^2$ drops off when we are near the origin. The last term on the r.h.s. is the potential of the homogeneous field. In equation (\ref{alphadyn}), we require that the potential dominates the dynamics over the inhomogeneities given by $\alpha$ and $f$. This requirement leads us to the relations near the origin
\begin{eqnarray}
\Phi_0 e^{\bar{\alpha}} > \frac{4\sqrt{\alpha''}}{m},& & \Phi_0 > \sqrt{\Phi_0\bar{f}+ \bar{f}^2} \nonumber \\
\mbox{and}~~~ \frac{\Phi_0}{\bar{f}'} e^{\bar{\alpha}} &>& \frac{\sqrt{3}}{m}. 
\end{eqnarray}
Reminding the reader that, near the origin, these conditions involve only constants. Defining the inhomogeneity physical scale as ${\cal{D}} \sim | \Phi_0 / \Phi' | e^{\alpha}$, we get, from the last inequality above, remembering that $\Phi'=f'$, that the size of the initial inhomogeneity must be
\begin{equation}
{\cal{D}} > \frac{\sqrt{3}}{m}.\label{calD}
\end{equation}

The mass of the field is related to the horizon scale $H^{-1}$ during inflation through the Friedmann equation for inflationary expansion
\begin{equation}
H^2 = \frac{1}{6}V(\Phi_0) ~~~ \rightarrow  ~~~  \frac{1}{m} \sim \frac{1}{\sqrt{12}}\Phi_0 H^{-1} \label{inversem}.
\end{equation}

Having fulfilled every requirement listed above, we are now able to conclude that, from the conditions for sufficient inflation and from the equations (\ref{calD}) and (\ref{inversem}), the initial perturbation must be larger than a few horizons, or more specifically
\begin{equation}
{\cal{D}}>11,83 H^{-1} .
\end{equation}

We recall here that this result is independent of the initial shape of the inhomogeneous inflaton field. The only requirements for this result are the initial conditions $\dot{\Phi}=\beta=\dot{\beta}=0$ at the hypersurface $\eta_0$. The condition for $\beta$ is consequently satisfied as, from the definition of ${\cal D}$ and (\ref{tau}),
\begin{equation}
\tau_\beta \sim \frac{{\cal D}}{\Phi_0}, 
\end{equation}
which gives us $\tau_\beta > 3,54 H^{-1}$.

\section{Conclusion}

In the framework of an inhomogeneous spherically symmetric geometry and 
massive non self-interacting scalar field, we have shown that in order 
to obtain enough inflation to solve some of the problems of the standard 
cosmological model, one needs a homogeneous region with at least $11.83$
horizons. This result was obtained without any assumption about the inital
spatial configuration of the scalar field, except for the fact that it should be null at spatial infinity, and hence it is a generalization of
the results obtained in Ref.~\cite{calzetta}, where it was assumed a particular initial exponential
spatial distribution for the scalar field. 
We had also pointed out that the 
initial geometric configuration may have a physical justification as long
as it satisfies the Weyl curvature hypothesis \cite{weyl}, and that the kinetic term of the scalar field should be negligible at the
initial hypersurface. Hence, the necessary conditions for the occurrence of
inflation in a spherically symmetric geometry with a massive free scalar field
under the Weyl curvature hypothesis are that there should be an homogeneous
region of about 12 horizons size, and that the kinetic term of the scalar field
should be zero.

Our results shows that the conclusions of Ref.~\cite{calzetta} are rather robust and they
are in accordance with other ressults \cite{piran}. Note that we have already assumed
a special symmetry for the fields, which is spherical symmetry. It is expected that in
the general case, without imposing any symmetry at all from the beginning, the situation
may become worst, although no calculation has been done in this framework.
Hence our results, together with the results of other works \cite{piran,calzetta}, indicates
that inflation is not sufficient to solve the homogeneity problem: a theory of
initial conditions is imperative in order to solve this deep issue of cosmology.

\section{Acknowledgments}

We would like to thank CNPq of Brazil for financial support. We very gratefully acknowledge
various enlightening conversations with J\'er\^ome Martin.

\end{document}